\documentclass[11pt,twoside]{article}

\usepackage{asp2006}
\usepackage{epsf}
\usepackage{graphicx}
\usepackage{psfig}
\usepackage{lscape}

\begin{document}
\title{Constraints on Secondary Eclipse Probabilities of Long-Period Exoplanets from Orbital Elements}   
\author{Kaspar von Braun \& Stephen R. Kane}   
\affil{NASA Exoplanet Science Institute, California Institute of Technology}    

\begin{abstract} 

Long-period transiting exoplanets provide an opportunity to study the mass-radius relation and internal structure of extrasolar planets. Their studies grant insights into planetary evolution akin to the Solar System planets, which, in contrast to hot Jupiters, are not constantly exposed to the intense radiation of their parent stars. Observations of secondary eclipses allow investigations of exoplanet temperatures and large-scale exo-atmospheric properties. In this short paper, we elaborate on, and calculate, probabilities of secondary eclipses for given orbital parameters, both in the presence and absence of detected primary transits, and tabulate these values for the forty planets with the highest primary transit probabilities. 
\end{abstract}


\section{Introduction} \label{introduction}  

Secondary eclipses of exoplanets provide unique insight into their astrophysical properties such as surface temperatures, atmospheric properties, and efficiency of energy redistribution. In \citet{kvb08, kvb09}, we demonstrate that the probability of detecting transits or eclipses among known radial-velocity (RV) planets is sensitively dependent on the values of orbital eccentricity $e$ and argument of periastron $\omega$. with some combinations' of $e$ and $\omega$ making transit/eclipse searches among long-period planets viable. Though it is feasible to detect planetary eclipses from space \citep[e.g., ][]{ldl09} and even from the ground \citep[e.g., ][]{slm09}, the difference in signal-to-noise ratio between transits and eclipses makes detections of the former much more straightforward. In this paper, we calculate the probability of planetary eclipses with or without the knowledge of the existence of a primary transit.

\section{Planetary Eclipse Probabilities} \label{probabilities}    

The {\it a priori} geometric eclipse probability of an extrasolar planet, $P_e$, is
\begin{equation}
  P_e = \frac{(R_{planet} + R_\star)(1 + e \cos (3\pi/2 - \omega))}{a (1 - e^2)},
  \label{eclprob}
\end{equation}
where $R_{planet}$ and $R_\star$ are planetary and stellar radii, respectively \citep{kvb09}. $P_e$ is highest for $\omega = 3\pi / 2$.

The presence of an observed transit imposes a lower limit to the orbital inclination angle $i$, which, using equations 8--11 in \citet{kvb09}, provides the following (conditional) lower limit of $P_e$:

\begin{equation}
  P_e^{\prime} \geq \frac{(R_\star + R_{planet})}{(R_\star - R_{planet})} \frac{(1 - e \cos \omega)}{(1 + e \cos \omega)}.
  \label{newprob}
\end{equation}


\section{Discussion} \label{discussion}

\begin{figure*}
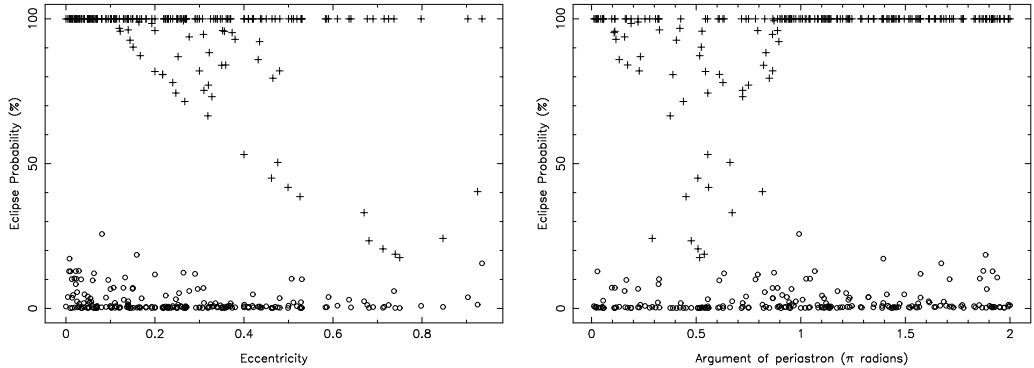

  \begin{center}
    \begin{tabular}{cc}
      \includegraphics[angle=270,width=6.6cm]{o_vonBraun_f1.ps} &
      \includegraphics[angle=270,width=6.6cm]{o_vonBraun_f2.ps} \\
    \end{tabular}
  \end{center}
  \caption{{\small The {\it a priori} values of $P_e$ (Eq. \ref{eclprob}; open circles) and conditional
    lower limits of $P_e$ (Eq. \ref{newprob}; crosses) for 203
    planets from the \citet{b06} catalog, plotted as a function of
    $e$ and $\omega$. For purpose of comparison, we assume solar and Jupiter masses and radii for all systems. }
}
  \label{newgep}
\end{figure*}

Fig. \ref{newgep} plots the {\it a priori} values (open circles) and conditional lower limits (crosses) of $P_e$ for 203 planets as functions of $e$ and $\omega$ values tabulated in \citet{b06}. The {\it left} panel clearly shows that, for low values of $e$, the existence of a transit practically guarantees an observable eclipse, whereas for higher eccentricities, this is not the case due to the correspondingly weaker constraint on inclination angle imposed by an existing transit \citep[cf. eq. 8 in ][]{kvb09}. The {\it right} panel demonstrates that, for $\omega \sim 3\pi / 2$, a detected transit ensures the existence of an observable eclipse , whereas for $\omega \sim \pi / 2$, this constraint is much weaker \citep[cf. fig. 1 in ][]{kvb09}. Thus, the presence of a planetary transit greatly affects the likelihood of existence of a secondary eclipse. Both the {\it a priori} value and the conditional lower limit of $P_e$ remain sensitively dependent on the combination of $e$ and $\omega$.

Table \ref{table1} shows the forty extrasolar planets with the (currently) highest transit probabilities, their orbital elements $P$, $e$, and $\omega$, and the explicit values for eclipse probabilities ($P_e$: {\it a priori} value; $P_e^{\prime}$: conditional lower limit). For instance, HD 118203 b has $P_e = 7.05\%$, but {\it if a primary transit is observed} ($P_t = 9.11\%$), then the existence of a secondary eclipse is almost certain ($P^\prime_e\geq94.65\%$). See table 1 in \citet{kvb09} for the equivalent inverse case of $P_t = f(P_e)$. 


\begin{table}[!ht] 
\caption{$P_t$, {\it a priori} $P_e$, and conditional (revised) $P^\prime_e$ for the 40 exoplanets from \citet{b06} with highest $P_t$ values. See \S\ref{discussion} \label{table1}} 
\smallskip 
\begin{center} 
{\small 
\begin{tabular}{lcccccc} 
\tableline 
\noalign{\smallskip} 
Planet & $P$ (days) & $e$ & $\omega (\deg)$ & $P_t$ (\%) & $P_e$ (\%) & $P^{\prime}_e$ (\%)\\ 
\noalign{\smallskip} 
\tableline 
\noalign{\smallskip} 
HD 41004 B b &      1.33 &  0.08 &  178.50 &   25.81 &   25.70 &  100.00 \\
HD 86081 b &      2.14 &  0.01 &  251.00 &   16.93 &   17.19 &  100.00 \\
GJ 436 b &      2.64 &  0.16 &  339.00 &   16.50 &   18.50 &  100.00 \\
55 Cnc e &      2.80 &  0.26 &  157.00 &   15.17 &   12.33 &   99.36 \\
GJ 674 b &      4.69 &  0.20 &  143.00 &   14.93 &   11.72 &   95.96 \\
HD 46375 b &      3.02 &  0.06 &  114.00 &   13.58 &   12.11 &  100.00 \\
HD 187123 b &      3.10 &  0.01 &    5.03 &   12.81 &   12.78 &  100.00 \\
HD 83443 b &      2.99 &  0.01 &  345.00 &   12.77 &   12.82 &  100.00 \\
HD 179949 b &      3.09 &  0.02 &  192.00 &   12.74 &   12.86 &  100.00 \\
HD 73256 b &      2.55 &  0.03 &  337.00 &   12.66 &   12.95 &  100.00 \\
HD 102195 b &      4.12 &  0.06 &  109.90 &   10.85 &    9.69 &  100.00 \\
HD 76700 b &      3.97 &  0.09 &   29.90 &   10.82 &    9.84 &  100.00 \\
HD 75289 b &      3.51 &  0.03 &  141.00 &   10.47 &   10.03 &  100.00 \\
51 Peg b &      4.23 &  0.01 &   58.00 &   10.35 &   10.12 &  100.00 \\
$\tau$ Boo b &      3.31 &  0.02 &  188.00 &   10.21 &   10.27 &  100.00 \\
HD 88133 b &      3.42 &  0.13 &  349.00 &   10.16 &   10.68 &  100.00 \\
BD -10 3166 b &      3.49 &  0.02 &  334.00 &   10.15 &   10.32 &  100.00 \\
HAT-P-2 b &      5.63 &  0.51 &  184.60 &    9.44 &   10.24 &  100.00 \\
HD 17156 b &     21.20 &  0.67 &  121.00 &    9.14 &    2.47 &   33.05 \\
HD 118203 b &      6.13 &  0.31 &  155.70 &    9.11 &    7.05 &   94.65 \\
$\upsilon$ And d &      4.62 &  0.02 &   57.60 &    8.69 &    8.38 &  100.00 \\
HD 68988 b &      6.28 &  0.15 &   40.00 &    8.20 &    6.76 &  100.00 \\
HIP 14810 b &      6.67 &  0.15 &  160.00 &    7.86 &    7.10 &  100.00 \\
HD 162020 b &      8.43 &  0.28 &   28.40 &    7.84 &    6.02 &   93.77 \\
HD 217107 b &      7.13 &  0.13 &   20.00 &    7.76 &    7.11 &  100.00 \\
HD 168746 b &      6.40 &  0.11 &   17.40 &    7.63 &    7.16 &  100.00 \\
HD 185269 b &      6.84 &  0.30 &  172.00 &    7.30 &    6.72 &  100.00 \\
HD 49674 b &      4.94 &  0.29 &  283.00 &    6.68 &   11.94 &  100.00 \\
HD 69830 b &      8.67 &  0.10 &  340.00 &    6.24 &    6.68 &  100.00 \\
HD 130322 b &     10.71 &  0.03 &  149.00 &    5.76 &    5.62 &  100.00 \\
HD 33283 b &     18.18 &  0.48 &  155.80 &    5.31 &    3.56 &   82.03 \\
HD 38529 b &     14.31 &  0.25 &  100.00 &    5.21 &    3.17 &   74.40 \\
HD 55 Cnc b &     14.65 &  0.02 &  164.00 &    4.67 &    4.63 &  100.00 \\
HD 13445 b &     15.76 &  0.04 &  269.00 &    4.47 &    4.85 &  100.00 \\
HD 27894 b &     17.99 &  0.05 &  132.90 &    4.43 &    4.12 &  100.00 \\
HD 108147 b &     10.90 &  0.53 &  308.00 &    4.14 &   10.09 &  100.00 \\
HD 6434 b &     22.00 &  0.17 &  156.00 &    4.02 &    3.50 &  100.00 \\
HD 190360 c &     17.11 &  0.00 &  168.00 &    3.94 &    3.93 &  100.00 \\
HD 20782 b &    585.86 &  0.93 &  147.00 &    3.92 &    1.29 &   40.33 \\
GJ 876 c &     30.34 &  0.22 &  198.30 &    3.85 &    4.44 &  100.00 \\
HD 99492 b &     17.04 &  0.25 &  219.00 &    3.83 &    5.29 &  100.00 \\
\noalign{\smallskip} 
\tableline 
\end{tabular} 
}
\end{center}
\end{table}


\end{document}